\documentclass{elsart}
\usepackage{amsmath}

\input{tcilatex}

\begin{document}

\begin{frontmatter}%

\title{An econophysics approach to analyse uncertainty in financial markets: an application to the Portuguese stock market}%

\author{*Andreia Dionisio, **Rui Menezes and **Diana A. Mendes}%

\address{*University of Evora, Center of Business Studies, CEFAG-UE, 
Largo Colegiais, 2, 7000 Evora, Portugal 
E-mail: andreia@uevora.pt
**ISCTE
Departament of Quantitative Methods, Av. Forcas Armadas, 1649-Lisboa, Portugal
E-mail: rui.menezes@iscte.pt, diana.mendes@iscte.pt}%

\begin{abstract}
In recent years there has been a closer interrelationship between several
scientific areas trying to obtain a more realistic and rich explanation of
the natural and social phenomena. Among these it should be emphasized the
increasing interrelationship between physics and financial theory. In this
field the analysis of uncertainty, which is crucial in financial analysis,
can be made using measures of physics statistics and information theory,
namely the Shannon entropy. One advantage of this approach is that the
entropy is a more general measure than the variance, since it accounts for
higher order moments of a probability distribution function. An empirical
application was made using data collected from the Portuguese Stock Market.%
\end{abstract}%


\end{frontmatter}%

\section*{Introduction}

The application of mathematical and physics models to finance goes back to
Bachelier in 1900, where it tests the hypothesis that stock prices follow a
random walk. However this simple version of the model did not account for
important characteristics of price variations, such as the occurrence of
crashes, nonlinear serial dependence, etc. Bachelier assumed that the price
variations follow a normal distribution, constant over time, and do not pay
attention to extreme events. However, the empirical evidence has shown that
stock prices seldomly behave in such a way as described by Bachelier.

The stock markets are usually complex systems, because they are open systems
where innumerous subsystems act and interact in a nonlinear and dynamic way,
constituting an attraction for the physicists that studied the working of
financial markets using different methods than those used by traditional
economists.

Bonanno, Lillo and Mantegna (2001) consider that the financial markets show
several levels of complexity that may occurred for being systems composed by
agents that interact nonlinearly between them. These authors, among others,
consider that the traditional models of asset pricing (CAPM and APT) failed
because the basic assumptions of these models are not verified empirically.

The entropy is a measure of dispersion, uncertainty, disorder and
diversification used in dynamic processes, in statistics and information
theory, and has been increasingly adopted in financial theory [Horowitz 
\emph{et al.} (1968), Philippatos\emph{\ et al.} (1972), Buchen \emph{et al.}
(1996), Zellner (1996), Molgedey\emph{\ et al}\textit{.} (2000), Stuzer
(2000), London \emph{et al.} (2001)].

In addition to the studies mentioned above, Bouchaud, Potters and Aguilar
(1997) have used entropy as an integrating measure in the process of
portfolio selection based on the mean-variance model of Markowitz. This is
because information is imperfect and the theoretical assumptions of
portfolio selection models do not apply in the reality. The authors suggest
the use of entropy with the purpose of obtaining a minimum diversification
and, at the same time, an acceptable risk level to the investor. In a
slightly different context, Fernholz (1999) and Samperi (1999) analysed the
entropy as a measure of diversification in financial markets. Gulko (1998)
analyses market equilibrium by building a model where entropy is maximized
subject to certain restrictions. He defends the \textquotedblleft entropy
pricing theory\textquotedblright\ as the main characteristic of market
efficiency.

The use of entropy as a measure of uncertainty in finance appears to have
many potentialities and a vast field of development, both in theoretical and
empirical work. In line with the above arguments, this paper examines the
ability of entropy as a measure of uncertainty in portfolio management
applied to the Portuguese stock market, highlighting the fact that entropy
verifies the effect of diversification.

In this article, the notion of uncertainty is related with the greater or
lesser difficulty to predict the future. Generally, it is normal to relate
the variance or the standard-deviation and the VaR (Value-at-Risk) as the
main risk and uncertainty measures in finance. However, some authors [e.g.
Soofi (1997), Maasoumi (1993), McCauley (2003)] alert for the fact that
these measures can fail in specific situations as a measure of uncertainty,
since they need that the probability distributions are symmetric and neglect
the possibility of extreme events such as the existence of fat-tails.

\section{Theoretical background}

Suppose that we have a set of possible events whose probabilities of
occurrence are $p_{1}$, $p_{2},...,p_{n}$ and $H$ is a measure of
uncertainty. According to Shannon (1948), a good measure of uncertainty $%
H=H(p_{1},...,p_{n})$ should satisfy the following properties:

\begin{enumerate}
\item $H$ should be continuous in $p_{i},i=1,...,n;$

\item If $p_{i}=1/n$, then $H$ should be a monotonic increasing function of $%
n$;

\item $H$ is maximized in a uniform probability distribution context;

\item $H$ should be additive;

\item $H$ should be the weighted sum of the individual values of $H$
\end{enumerate}

According to Shannon (1948) one measure that satisfies all these properties
is the entropy which is defined as $H\left( X\right) =-\underset{i}{\sum }%
p_{i}\log p_{i}$. When the random variable has a continuous distribution,
and $p_{X}(x)$ is the density function of the random variable $X$, the
entropy is given by%
\begin{equation}
H\left( X\right) =-\int p_{X}(x)\log p_{X}(x)dx.  \label{entropy-cont}
\end{equation}

The properties of the entropy of continuous and discrete distributions are
mainly alike. In particular we have [Shannon (1948); Kraskov\emph{\ et al.}
(2004)]:

\begin{description}
\item[$(a)$] If $X$ is limited to a certain volume $v$ in its space, then $%
H(X)$ is a maximum and is equal to $\log v$ when $p_{X}(x)$ is constant, $%
1/v $, in the volume;

\item[$(b)$] For any two variables $X$ and $Y$, we have $H\left( X,Y\right)
\leq H\left( X\right) +H\left( Y\right) $ where the equality holds if (and
only if) $X$ and $Y$ are statistically independent, i.e. $%
p_{X,Y}(x,y)=p_{X}(x)p_{Y}(y);$

\item[$(c)$] The joint entropy can be given by $H\left( X,Y\right) =H\left(
X\right) +H\left( Y|X\right) =H\left( Y\right) +H\left( X|Y\right) ,$since $%
H\left( X\right) +H\left( Y\right) \geq H\left( X,Y\right) ,$ then $H\left(
Y\right) \geq H\left( Y|X\right) $ and $H\left( X\right) \geq H\left(
X|Y\right) .$
\end{description}

The assumption that the data and the residuals follow a normal distribution
is very common in portfolio management and regression analysis. Thus, the
equation used to estimate parametrically the entropy of a normal
distribution, $NH\left( X\right) $, is%
\begin{equation}
NH\left( X\right) =\int p_{X}(x)\log \sqrt{2\pi }\sigma dx+\int p_{X}(x)%
\frac{\left( x-\overset{\_}{x}\right) ^{2}}{2\sigma ^{2}}dx=\log \left( 
\sqrt{2\pi e}\sigma \right) .  \label{normal}
\end{equation}

Arafat, Skubic and Keegan (2003) consider that a measure of uncertainty
should attend to the following properties: $\left( i\right) $ \emph{Symmetry}%
, that is $H\left( X\right) =H\left( 1-X\right) ;$ and $\left( ii\right) $ 
\emph{Valuation}: $H\left( X\cup Y\right) +H\left( X\cap Y\right) =H\left(
X\right) +H\left( Y\right) .$ These authors discuss combined methods of
uncertainty and conclude that entropy can be a good measure of uncertainty.

One of the difficulties to estimate the mutual information on the basis of
empirical data lies on the fact that the underlying \emph{pdf }is unknown.
To overcome this problem, there are essentially three different methods to
estimate mutual information: \ histogram-based estimators,\ kernel-based
estimators and\ parametric methods.\footnote{%
The histogram-based estimators are divided in two groups: equidistant cells
and equiprobable cells, i.e. marginal equiquantisation [see e.g.Darbellay
(1998)]. The second approach presents some advantages, since it allows for a
better adequacy to the data and maximizes mutual information [Darbellay
(1998)].} In order to minimize the bias that may occur, we will use the
marginal equiquantization estimation process, proposed by Darbellay (1998).

The introduction of entropy as a measure of uncertainty in finance goes back
to Philippatos and Wilson (1972), which present a comparative analysis
between the behaviour of the standard-deviation and the entropy on portfolio
management. These authors conclude that entropy is more general and has some
advantages facing to the standard-deviation. According to Lawrence (1999)
the two main measures of uncertainty are entropy and variance, because
entropy is a concave function allows its use as an uncertainty function.

\section{Entropy and diversification effect: an example}

Historically, the variance has had a fundamental role in the analysis of
risk and uncertainty. However, according to Maasoumi (1993), entropy can be
an alternative measure of dispersion and in addition Soofi (1997) considers
that the interpretation of the variance as a measure of uncertainty must be
done with some precaution.

The entropy is a measure of disparity of the density $p_{X}\left( x\right) $
from the uniform distribution. It measures uncertainty in the sense of
"utility" of using $p_{X}\left( x\right) $ in place of the uniform
distribution. The variance measures an average of distances of outcomes of
the probability distribution from the mean. According to Ebrahimi, Maasoumi
and Soofi (1999), both measures reflect concentration but their respective
metrics of concentration are different. Unlike the variance that measures
concentration only around the mean, the entropy measures diffuseness of the
density irrespective of the location of concentration.

In terms of mathematical properties, entropy $\left[ H\left( X\right) \right]
$ is non-negative in the discrete case. In the discrete case, $H\left(
X\right) $ is invariant under one-to-one transformations of $X$, but the
variance is not. For the continuous case, neither the entropy nor the
variance are invariant under one-to-one transformations. The entropy of a
continuous random variable $X$ takes values in $\left] -\infty ,+\infty %
\right[ $ [Shannon (1948)].

Ebrahimi, Maasoumi and Soofi (1999) examined the role of \ variance and
entropy in ordering distributions and random prospects, and conclude that
there is no universal relationship between these measures in terms of
ordering distributions. These authors found that, under certain conditions,
the order of the variance and entropy is similar when continuous variables
are transformed and show (using a Legendre series expansion) that the
entropy depends on many more parameters of a distribution than the variance.
A Legendre series expansion reveals that entropy may be related to
higher-order moments of a distribution which, unlike the variance, could
offer a much better characterization of $p_{X}\left( x\right) $ since it
uses more information about the probability distribution than the variance.

In this paper we examine the sensitivity of entropy to the effect of
diversification. The risk of a portfolio can be splitted into specific risk
and systematic risk, that is not diversifiable (see Figure \ref{Risk}).
Using entropy we can obtain a similar type of information, since $H\left(
X\right) =I\left( X,Y\right) +H\left( X|Y\right) ,$ where $I\left( .\right) $
is the mutual information between $X$ and $Y$ and may be comparable with the
systematic risk and $H\left( .|.\right) $ is the conditional entropy that
can be comparable with the specific risk. We must emphasize that the
measures of information theory are not directly comparable to the analysis
of variance in metric terms.

\FRAME{ftbphFU}{4.6674in}{2.7008in}{0pt}{\Qcb{Specific risk and systematic
risk.}}{\Qlb{Risk}}{dionisiofigure1.eps}{\special{language "Scientific
Word";type "GRAPHIC";maintain-aspect-ratio TRUE;display "USEDEF";valid_file
"F";width 4.6674in;height 2.7008in;depth 0pt;original-width
6.1272in;original-height 3.5336in;cropleft "0";croptop "1";cropright
"1";cropbottom "0";filename '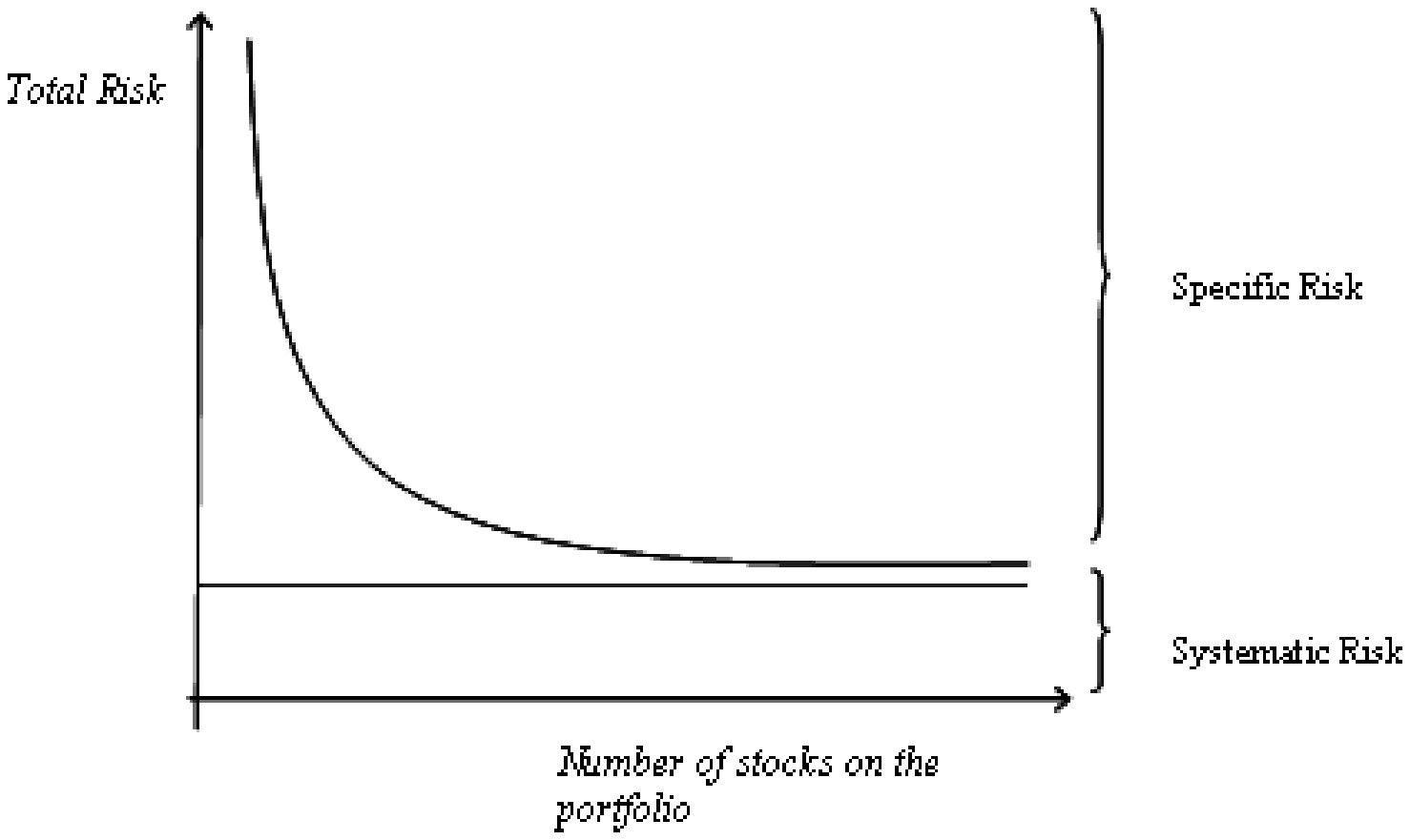';file-properties "XNPEU";}}

It is important to note some properties of the variance (and
standard-deviation) and entropy as measures of uncertainty. The
standard-deviation is a convex function, which according to the Jensen
inequality $E\left[ \sigma \left( X\right) \right] \geq \sigma \left[ \left(
EX\right) \right] .$\footnote{%
The equality occurs when the linear correlation coefficient between the
variables is 1.} This property allows the variance and the
standard-deviation to be used as risk measures in stock portfolios, since
they take into account the effect of diversification.

The entropy is a concave function and has a maximum for most of the
probability distributions, and this fact leads us to think that entropy will
not satisfy the effect of diversification. However, we must note that
entropy is not a function of the values of the variables but the probability
itself and the property $H\left( X,Y\right) \leq H\left( X\right) +H\left(
Y\right) $ can bring some hope in this way.

In this paper we perform a similar analysis to that presented by Elton and
Gruber (1995). These authors showed that diversification is a factor of
minimization of the specific risk (measured by the standard-deviation). They
made a random selection of the assets to compose portfolios and the only
premise is the fact that the proportion invested in each asset is $1/N$,
being $N$ the number of assets in the portfolio. We use daily closing prices
of 23 stocks rated on the Portuguese stock market (\emph{Euronext Lisbon}),
spanning from 28/06/1995 to 30/12/2002, which corresponds to 1856
observations \emph{per} stock, in order to compute the rates of return. The
statistical analysis of these time series revealed that we must reject the
null that the empirical distributions are normal, since they show high
levels of kurtosis and skewness.

In order to compare the behaviour of entropy with the standard-deviation in
a coherent way, we use the normal entropy (equation \ref{normal}), since the
normal entropy is a function of the standard-deviation.

Our results (see Figure \ref{comparative}) show that the entropy and the
standard-deviation tend to decrease when we include one more asset in the
portfolio. This fact allows us to conclude that entropy is sensitive to the
effect of diversification. These results can be explained by the fact that
when the number of assets in the portfolio increases, the number of possible
states of the system (portfolio) declines progressively and the uncertainty
about that portfolio tends to fall. Besides, we verify that the entropy
respects the condition of subadditivity suggested by Reesor and McLeish
(2002), where $H[\theta X]+H[(1-\theta )Y]\geq H[\theta X+(1-\theta )Y]$,
being $\theta =1/N.$

\FRAME{ftbpFU}{4.4901in}{2.674in}{0pt}{\Qcb{Comparative analysis of the
empirical entropy ($H$) and the normal entropy ($NH$) for portfolios
randomly selected. Entropy is measured in $nats$ because we use natural
logarithms.}}{\Qlb{comparative}}{dionisofigure2.eps}{\special{language
"Scientific Word";type "GRAPHIC";maintain-aspect-ratio TRUE;display
"USEDEF";valid_file "F";width 4.4901in;height 2.674in;depth
0pt;original-width 6.2111in;original-height 3.6867in;cropleft "0";croptop
"1";cropright "1";cropbottom "0";filename
'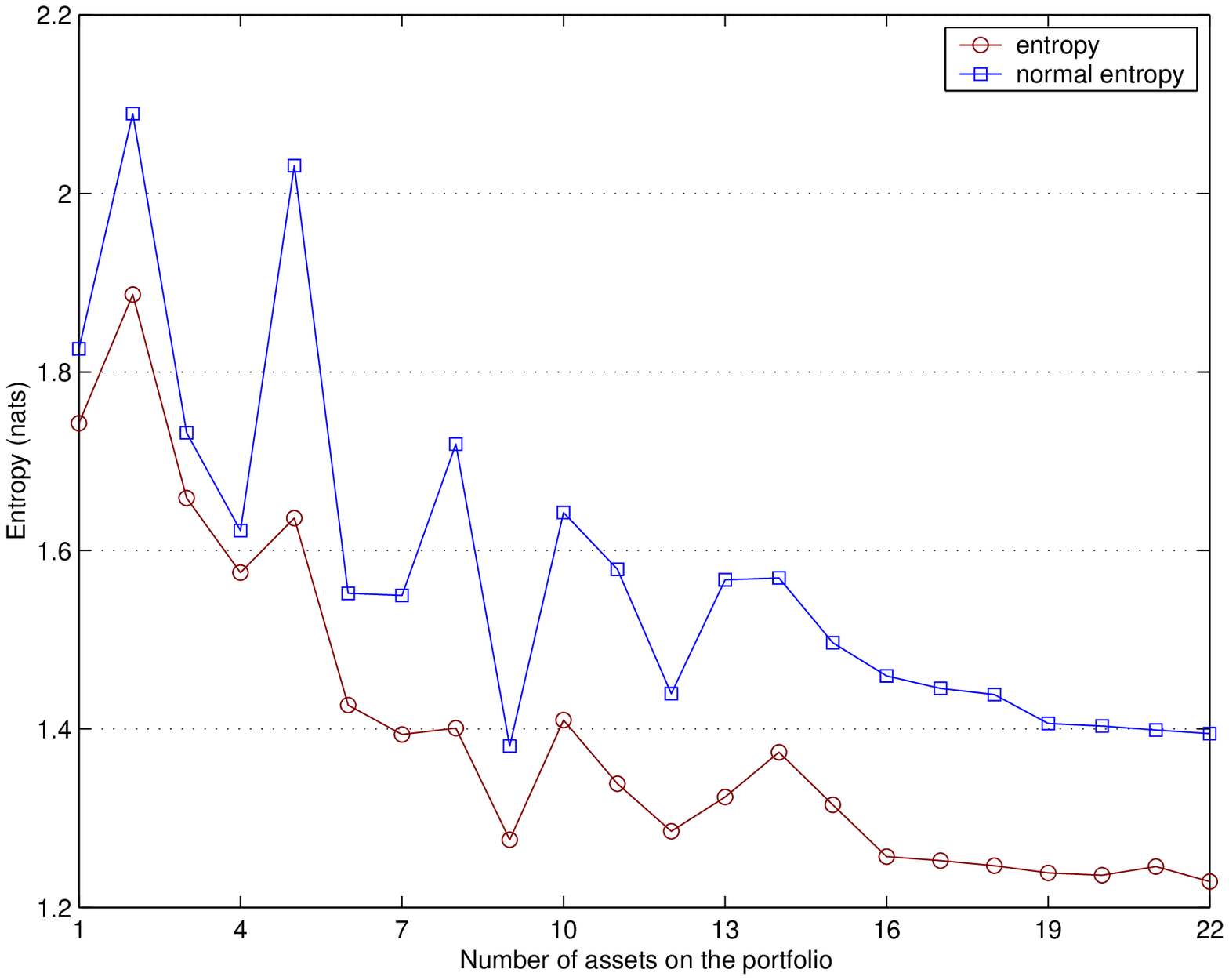';file-properties "XNPEU";}}

We must highlight the fact that, in this example, the normal entropy assumes
always higher values than the empirical entropy. This means that the
predictability level of each portfolio is higher than the one assumed by the
normal distribution.

From this preliminary analysis, we can conclude that entropy observes the
effect of diversification and is a more general uncertainty measure than the
variance, since it uses much more information about the probability
distribution.

\section{Final remarks}

This paper analyses the use of entropy as a measure of uncertainty in
portfolio management. This can be a complementary way to the traditional
mean-variance models, whose assumptions are typically quite restrictive. Our
approach takes into account the higher-order moments of the empirical
probability distributions, rather than just the variance that only uses the
second moment.

The results suggest that entropy is sensitive to the effect of
diversification and is apparently a more general measure of uncertainty than
the variance.

\bigskip

\textbf{Acknowledgement}

Financial support from the Funda\c{c}\~{a}o Ci\^{e}ncia e Tecnologia,
Lisbon, is grateful acknowledged by the third author, under the contract N.%
${{}^o}$
POCTI/ ECO/ 48628/ 2002, partially funded by the European Regional
Development Fund (ERDF)."

\end{document}